\documentclass[%
reprint,
superscriptaddress,
amsmath,amssymb,
aps,
prb,
]{revtex4-1}

\usepackage{amssymb}
\usepackage{fullpage}
\usepackage{graphicx}
\usepackage{bibunits}
\usepackage{wrapfig}
\usepackage{floatflt}
\usepackage{color}
\usepackage{xspace}
\usepackage{dsfont}
\usepackage[official,right]{eurosym}
\usepackage[top=0.8in, bottom=0.75in, left=0.75in, right=0.75in]{geometry}

\usepackage{graphicx}
\usepackage{dcolumn}
\usepackage{bm}
\usepackage{hyperref}

\usepackage{multirow}

\newcommand{\BPCB} {(C$_5$H$_{12}$N)$_2$CuBr$_4$\xspace}

\newcommand{\BaCuGeO}{Ba$_2$CuGe$_2$O$_7$\xspace}

\newcommand{\ket}[1]{\left| #1 \right\rangle}

\newcommand{\be}{\begin{equation} }
\newcommand{\ee}{\end{equation} }
\newcommand{\bea}{\begin{eqnarray} }
\newcommand{\eea}{\end{eqnarray} }

\newcommand{\mb}[1]{\mathbf{#1}}
\newcommand{\mr}[1]{\mathrm{#1}}
\newcommand{\mc}[1]{\mathcal{#1}}
\newcommand{\al}{\alpha}

\def\XXint#1#2#3{{\setbox0=\hbox{$#1{#2#3}{\int}$}
     \vcenter{\hbox{$#2#3$}}\kern-.5\wd0}}

\begin{document}

\title{Origin of magnetic anisotropy in the spin ladder compound \BPCB}

\author{D. Blosser}
\email{dblosser@phys.ethz.ch}
\affiliation{Laboratory for Solid State Physics, ETH Z\"urich, 8093 Z\"urich, Switzerland}

\author{V. K. Bhartiya}
\affiliation{Laboratory for Solid State Physics, ETH Z\"urich, 8093 Z\"urich, Switzerland}

\author{D. J. Voneshen}
\affiliation{ISIS Facility, Rutherford Appleton Laboratory, Chilton, Didcot, Oxon OX11 0QX, United Kingdom}

\author{A. Zheludev}
\email{zhelud@ethz.ch}
\homepage{http://www.neutron.ethz.ch/}
\affiliation{Laboratory for Solid State Physics, ETH Z\"urich, 8093 Z\"urich, Switzerland}

\date{\today}

\begin{abstract}

The $S=1/2$ spin ladder compound \BPCB (BPCB) is studied by means of high-resolution inelastic neutron scattering. In agreement with previous studies we find a band of triplet excitations with a spin gap of $\sim0.8$~meV and a bandwidth of $\sim0.6$~meV. In addition, we observe a distinct splitting of the triplet band of $50(1)$~$\mu$eV or $40(2)$~$\mu$eV at the band minimum or maximum, respectively. 
By comparison to a strong coupling expansion calculation of the triplet dispersion for a spin ladder with anisotropic exchange, weakly anisotropic leg interactions are identified as the dominant source of magnetic anisotropy in BPCB. 
Based on these results, we discuss the nature of magnetic exchange anisotropy in BPCB and in related transition-metal insulators. 

\end{abstract}


\maketitle

\section{Introduction}
The Heisenberg model is arguably the most important construct in quantum magnetism. Experimentally, it is usually studied in transition metal oxides and metal-organic salts.
Unfortunately, symmetry-breaking magnetic anisotropy is unavoidable in such compounds. Even when weak compared to the dominant Heisenberg exchange, it can have a significant influence on the magnetic and dynamical properties.
In understanding these, existing microscopic theories are
usually disregarded in favor of minimalistic empirical models. 
For example, magnetic exchange interactions in insulators are described within Anderson's theory of super exchange\cite{Anderson1959,Moriya1960PRL,Moriya1960}. In the presence of weak spin-orbit-coupling it predicts both symmetric (Ising) and antisymmetric (Dzyaloshinskii-Moriya, DM) exchange anisotropy always occurring in combination and at a fixed ratio\cite{Kaplan1983,SEWA1992,SEWA1993}. 
Yet, experimental data is often successfully analyzed in comparison to a Heisenberg model with only an added DM term  \cite{Oshikawa1997,Affleck1999,Zvyagin2004,Coldea2002,Haelg2014} even though such a model lacks microscopic justification.
Even more often, signatures of weak magnetic anisotropy are detected in experiments, but their origin remains unknown altogether \cite{Kenzelmann2001,Jacobsen2018,Bettler2019}.
In short, anisotropy is ubiquitously present, yet pinpointing its microscopic origin is almost never possible.

Here we present a detailed case study: the effect of very weak magnetic anisotropy on spin excitations in
the almost ideal Heisenberg quantum spin ladder compound \BPCB{} (BPCB)\cite{Patyal1990,Savici2009,Ruegg2008BPCBThermodynamics,Klanjsec2008NMROrdering,ThielemannRuegg2009,CizmarBPCBAniso2010,Blosser2018}. 
On this compound we have obtained exceptionally high resolution inelastic neutron scattering data showing a small but distinct anisotropy splitting of the triplet excitations as described in Sec.~\ref{sec:Experiment}. 
At the same time, BPCB is an excellent realization of a strong rung spin ladder. It is described by only two exchange couplings: the rung exchange and an almost four times smaller leg coupling. For this model, the excitation spectrum in the presence of various possible exchange anisotropies can be precisely calculated as detailed in Sec.~\ref{sec:TripletDispersion}.
This combination of high resolution spectroscopic data and precise calculations of the excitation spectrum allows a detailed discussion of the anisotropy's microscopic origins in BPCB (Sec.~\ref{sec:Discussion}). These findings we compare to the microscopic theory of anisotropic superexchange and we discuss how the present case study may guide our understanding of magnetic anisotropy in related compounds.

\section{Experiment and results}\label{sec:Experiment}

\begin{figure}
\includegraphics{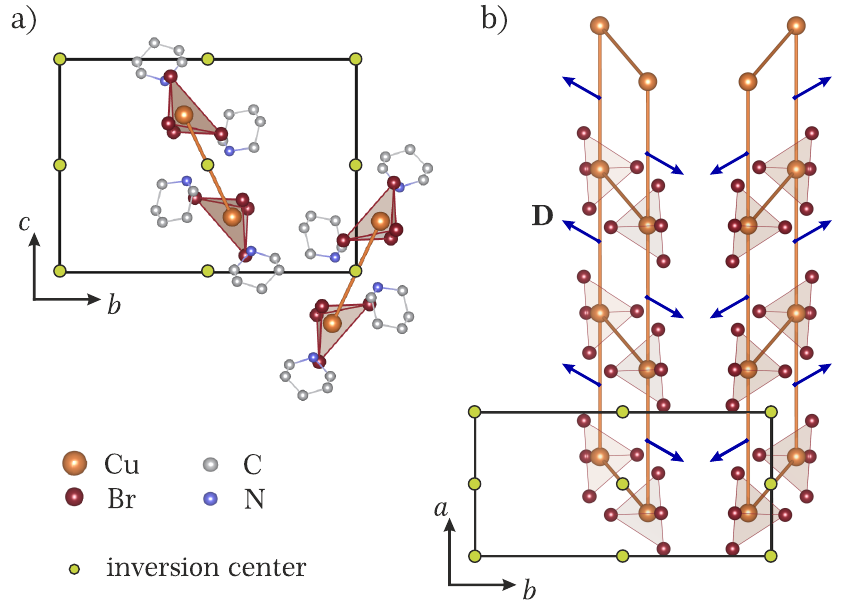}
\caption{\label{fig:structure} 
Schematic of the crystal structure of BPCB. The spin ladders are formed by the magnetic Cu${^{2+}}$ cations and linking Br${^{-}}$ anions. Crystallographic centers of inversion symmetry are marked by green dots. Anisotropic Dzyaloshinskii-Moriya interactions, by symmetry are only allowed on the ladder legs and possible DM vectors $\mb{D}$ are sketched as blue arrows on these bonds. 
Note that for this monoclinic structure, in both plots the third axis (projection axis) is not perpendicular to the plane of the figure.}
\end{figure}

\begin{figure*}[t]
\includegraphics{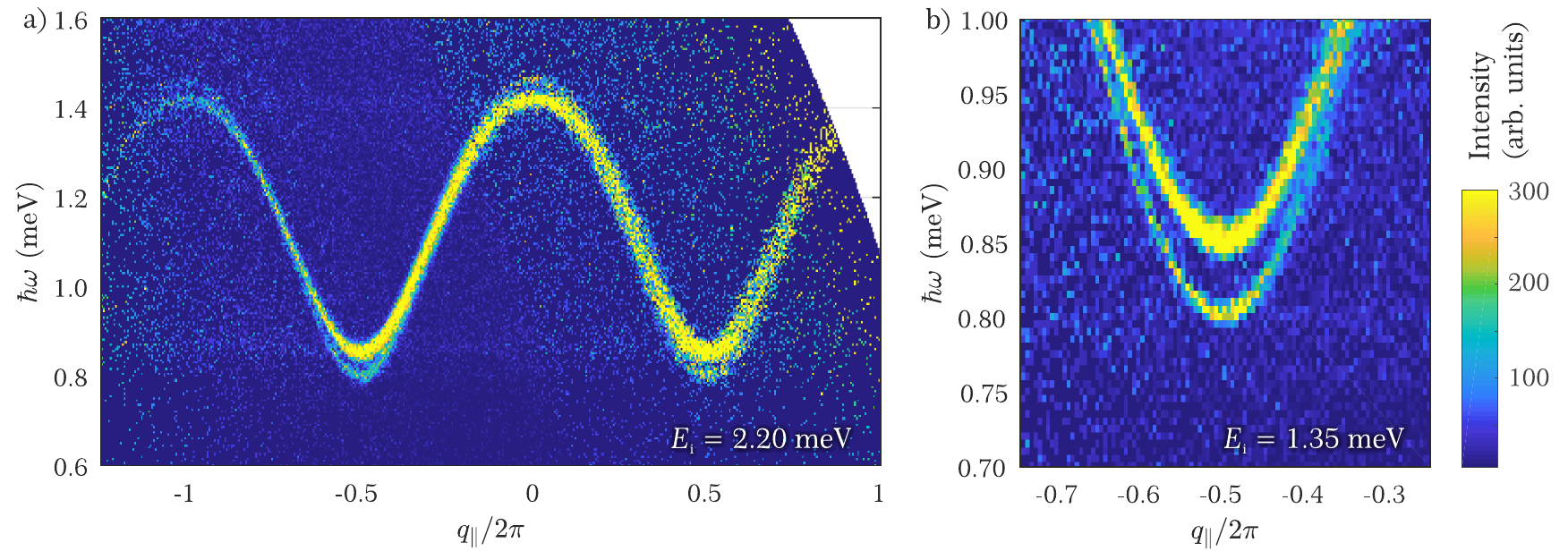}
\caption{\label{fig:spectra}False-color maps of the inelastic neutron scattering intensity measured using a) $E_\mr{i}=2.20$~meV and b) $E_\mr{i}=1.35$~meV incident energy neutrons at $T=0.35$~K. With the lower incident energy, a superior energy resolution is achieved but only the low energy parts of the dispersion can be probed.}
\end{figure*}

\subsection{The compound \BPCB: crystal structure and magnetic interaction pathways}
The compound \BPCB, bis-piperidinium copper bromide or BPCB for short crystallizes in a monoclinic crystal structure with space group P2$_1$/c and room temperature lattice parameters $a=8.49$, $b=17.22$, $c=12.38$~\AA{}, and $\beta=99.3^{\circ}$ \cite{Patyal1990}.
The spin ladders are formed by the magnetic Cu$^{2+}$ cations carrying a spin $S=1/2$ linked by super-exchange bridges via Br$^-$ anions \cite{Savici2009}.
The ladders run along the crystallographic $a$ axis and are well separated by non-magnetic organic piperidinium ions as depicted in Fig.~\ref{fig:structure}\footnote{The crystal structure plots shown in Fig.~\ref{fig:structure} are based on renderings generated by the VESTA software package\cite{VESTA2011}.} 

There are two crystallographically equivalent spin ladders in BPCB, related by the glide plane symmetry of the P2$_1$/c space group. Furthermore, the center of each ladder rung corresponds to a crystallographic center of inversion symmetry. For this reason, on the ladder rungs, only symmetric exchange is possible\cite{DZYALOSHINSKY1958}, whilst for the ladder legs there are no symmetry restrictions and both symmetric and antisymmetric exchange are in principle allowed. On the ladder leg, the antisymmetric exchange contribution is parameterized by a Dzyaloshinskii-Moriya vector $\mb{D}$. These DM vectors are uniform within every ladder leg and anti-aligned between the two legs of each ladder. The DM vectors of the two crystallographically equivalent ladders are again related by the reflection of the glide plane symmetry as sketched in Fig.~\ref{fig:structure}. Besides these relations, the DM vectors may point in any direction and are not further constrained by symmetry. 
Similarly, The $g$-tensors for the magnetic moments residing on the different Cu$^{2+}$ ions show different principal axes for the two types of ladders. 
Thus, in an applied magnetic field, the two types of ladders become inequivalent, except for special orientations of the magnetic field \cite{ThielemannRuegg2009,CizmarBPCBAniso2010}. In the present study, however, we exclusively focus on zero-field properties where the two ladders remain equivalent.

\subsection{Inelastic neutron scattering}

Neutron scattering experiments were performed on four fully deuterated single crystals of total mass 2.07~g, co-aligned to better than 1$^\circ$ effective mosaic spread. 
The measurements were carried out at the LET cold neutron time-of-flight spectrometer \cite{Bewley2011} at the ISIS facility, UK. 
The sample was mounted on a $^3$He-$^4$He dilution refrigerator 
with the crystallographic $b$ axis vertical.
Making use of repetition rate multiplication, data was collected simultaneously using neutrons of incident energies $E_\mr{i}=$ 1.35, 2.20, 4.20 and 11.0~meV. For these configurations we find an approximately Gaussian energy resolution of 19, 36, 97 and 410~$\mu$eV at FWHM, respectively, for elastic scattering ($\hbar\omega=0$) and improving towards higher energy transfer.


\begin{figure}
	\includegraphics{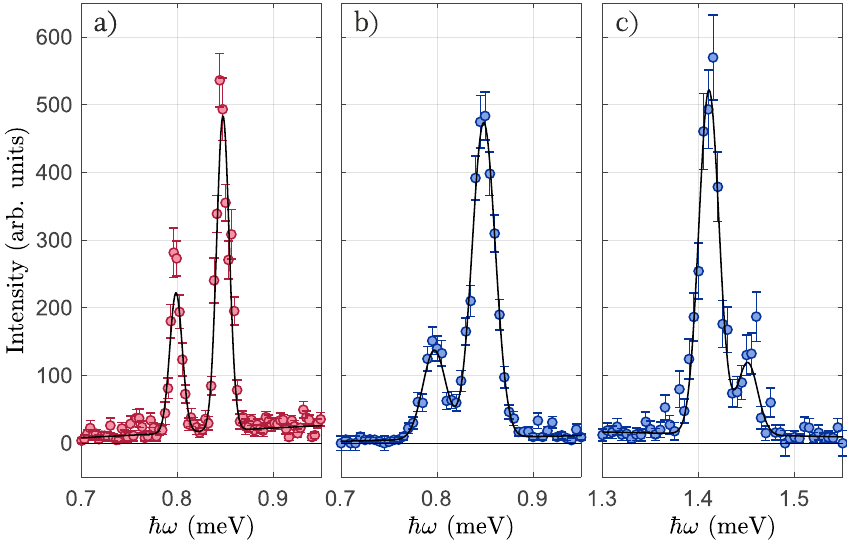}
	\caption{\label{fig:cuts} 
		Cuts through the data shown in Fig.~\ref{fig:spectra} at the band extrema: a) Cut at the band minimum $q_\|/2\pi=-0.5$ through the $1.35$~meV data set. b,c) Cuts at the band extrema $q_\|/2\pi=-0.5$ or $q_\|/2\pi=0$, respectively, through the $2.20$~meV data set.
		Solid lines correspond to Gaussian fits. At the band minimum we find a slightly larger splitting of $\Delta_\mr{min}=50(1)$~$\mu$eV than at the band maximum where $\Delta_\mr{max}=40(2)$~$\mu$eV.			}
\end{figure}

\subsection{Experimental results}

An overview of the neutron scattering data\footnote{The raw data is available in the STFC Research Data Repository \cite{DataISIS_BPCB}.}
collected for $E_\mathrm{i}=2.2$ and 1.35 meV at a temperature of $0.35$~K is presented as false color intensity plots in Fig.~\ref{fig:spectra}.
For this quasi-one-dimensional spin ladder system, the measured scattering intensity is plotted versus reduced wave-vector transfer parallel to the ladder $q_\|=\mb{Q}\cdot\mb{a}$ and integrated fully along the non-dispersive $b^*$ and $c^*$ directions.

We observe a band of excitations with an excitation gap of $\Delta\sim 0.8$~meV and a bandwidth of $\Gamma\sim 0.6$~meV. The line-width of gapped excitations in 1D systems is known to show activation behavior\cite{Zheludev2008}. With $k_\mr{B}T\ll\Delta$, in the present experiment these excitations are long lived and their apparent width corresponds to the experimental resolution.
Cuts through both data sets at the band extrema are shown in Fig.~\ref{fig:cuts}. These correspond to the data of Fig.~\ref{fig:spectra} integrated in thin slices of $q_\|/2\pi=-0.5\pm 0.02$ and $q_\|/2\pi=0\pm 0.03$, respectively (the dispersion is less curved at the maximum allowing to integrate a wider slice for improved counting statistics). 
To these cuts we fit two Gaussian peaks of equal width and a linear background shown as a solid lines in Fig.~\ref{fig:cuts}. We find a splitting of $\Delta_\mr{min}=50(1)$~$\mu$eV at the band minimum and $\Delta_\mr{max}=40(2)$~$\mu$eV at the band maximum.

\begin{widetext}
\section{Triplet dispersion for a spin ladder with exchange anisotropy}
\label{sec:TripletDispersion}

\subsection{Isotropic Heisenberg ladder} 
\label{sec:LadderHamiltonian_Heis}
For an isotropic spin $S=1/2$ ladder we denote the rung and leg Heisenberg exchange constants as $J_\bot$ and $J_\|$, respectively. In the limit of $\lambda=J_\|/J_\bot\ll1$, the spin ladder can be understood as a 1D array of weakly interacting spin dimers. For $\lambda=0$ the groundstate corresponds to a product state of singlets and the low energy excitations are triplet excitations. In the presence of weak leg coupling, these excitations become mobile and for small $\lambda$ their dispersion can be calculated in a strong coupling expansion\cite{ReigrotzkiRice1994}. 
For an isotropic ladder with Hamiltonian
\begin{equation}
\mc{H} = J_\bot \left( \mc{H}^\bot + \lambda \mc{H}^\| \right), \qquad
\mc{H}^\bot = \sum_{R=1}^{L}  \mb{S}_{R,1} \cdot \mb{S}_{R,2}, \qquad
\mc{H}^\| = \sum_{R=1}^{L} \sum_{\al=1}^{2}  \mb{S}_{R,\al} \cdot \mb{S}_{R+1,\al}, \qquad
\end{equation}
up to 3\textsuperscript{rd} order in $\lambda$, the following dispersion is obtained in Ref.~\onlinecite{ReigrotzkiRice1994},
degenerate for the three $\sigma=\{+,0,-\}$ triplet branches:
\begin{equation}
\epsilon_\sigma(k)/J_\bot =   1 + \gamma \cos(k) 
+ \frac{\gamma^2}{4} \Big[3-\cos(2k)\Big]    
+ \frac{\gamma^3}{8}\Big[3-2\cos(k)-2\cos(2k)+\cos(3k) \Big].
\label{eq:bpcb_TripletDispersion}
\end{equation}
{In the following we do the same calculation for a ladder with anisotropic rung or leg exchange, respectively. Some details of these computations are given in the appendix.}

\subsection{Anisotropic rung interactions} 
\label{sec:LadderHamiltonian_RungAniso}
In addition to the isotropic rung and leg couplings, here Ising-type anisotropy on the ladder rungs is considered. We choose the $z$ axis parallel the Ising axis and consider the Hamiltonian
\begin{equation}
\mc{H} = J_\bot \left( \mc{H}^\bot + \lambda \mc{H}' \right), \qquad
\mc{H}' = \mc{H}^\| + C\, \mc{H}^\mr{Rung,Ising}, \qquad
\mc{H}^\text{Rung,Ising} = \sum_{R=1}^{L}  {S}_{R,1}^z {S}_{R,2}^z,
\end{equation}
where the Ising anisotropy on the ladder rungs is parameterized by $C$. Using the same strong coupling expansion as applied to the isotropic case in Ref.~\onlinecite{ReigrotzkiRice1994}, up to 3\textsuperscript{rd} order in $\lambda$ we find
\begin{align}
	\epsilon_0(k)/J_\bot &= 
	1+ \lambda \cos(k)
	+\frac{\lambda^2}{4} \Big[3-\cos(2k)\Big]
	+\frac{\lambda^3}{8} \Big[3-2C  -2\cos(k)  -2\cos(2k)+\cos(3k)\Big]\\
	\epsilon_\pm(k)/J_\bot &=
	1 +\lambda \Big[\frac{C}{2}+\cos(k)\Big]
	+\frac{\lambda^2}{4} \Big[3-\cos(2k)\Big]
	+\frac{\lambda^3}{8} \Big[3 -2C -2\cos(k) +\big(C-2\big)\cos(2k)+\cos(3k)\Big].
\end{align}

\subsection{Anisotropic leg interactions}
\label{sec:LadderHamiltonian_LegAniso}
Finally, we consider anisotropic exchange on the ladder legs with both symmetric and anti-symmetric contributions. Assuming a center of inversion symmetry on the ladder rungs, we consider a uniform DM vector $+\mb{D}$ on one leg and $-\mb{D}$ on the other leg. Furthermore we include Ising anisotropy of independent magnitude but pointing in the same direction as $\mb{D}$. We choose the $z$ axis parallel to $\mb{D}$ and consider
\begin{align}\label{eq:LadderHamiltonian_LegAniso}
	\mc{H} = J_\bot \left( \mc{H}^\bot + \lambda \mc{H}' \right), \quad\quad
	\mc{H}' = \mc{H}^\| + &B\, \mc{H}^\mr{Leg,DM} + A\, \mc{H}^\mr{Leg,Ising}\\
	\mc{H}^\text{Leg,DM} = \sum_{R=1}^{L} \sum_{\al=1}^{2}  (-1)^\al \mb{e}_z\cdot(\mb{S}_{R,\al} \times \mb{S}_{R+1,\al}), \quad&\quad
	\mc{H}^\text{Leg,Ising} =\sum_{R=1}^{L} \sum_{\al=1}^{2}  
	S_{R,\al}^z S_{R+1,\al}^z
\end{align}
Up to 3\textsuperscript{rd} order in $\lambda$ we find the following triplet dispersion:
\begin{align}
	\epsilon_0(k)/J_\bot =& 
	1 + \lambda (1+A)\cos(k)  
	+ 
	\frac{\lambda^2}{4} \Big[3 -4B^2 + 2A + A^2 + 4B^2 \cos(k) - \Big(1 + 2A + A^2\Big)\cos(2 k) \Big] \notag\\
	&+ 
	\frac{\lambda^3}{8} \Big[3 +3A - \Big(2 +4B^2 +4A +4AB^2 +3A^2 + A^3\Big)\cos(k)   \notag\\&
	  \qquad\quad - 
	\Big(2 +2A -4B^2 -4AB^2 \Big)\cos(2k) 
	+ \Big(1 + 3A + 3A^2 + A^3\Big) \cos(3k) \Big]   \\ 
	\epsilon_\pm(k)/J_\bot =& 1 + \lambda \cos(k)
	+ 
	\frac{\lambda^2}{4} \Big[3 -2B^2 +2A + A^2 - \cos(2k) \Big] \notag\\
	&+ 
	\frac{\lambda^3}{8} \Big[3 +3A - \Big(2 +12B^2 +A +\frac{A^2}{2} \Big)\cos(k) 
	- 
	\Big(2 + 2B^2 +2A +2AB^2\Big)\cos(2k) 
	+ \frac{1}{8}\cos(3k) \Big]   
\end{align}

\end{widetext}

\section{Discussion}\label{sec:Discussion}

\begin{figure*}
	\includegraphics{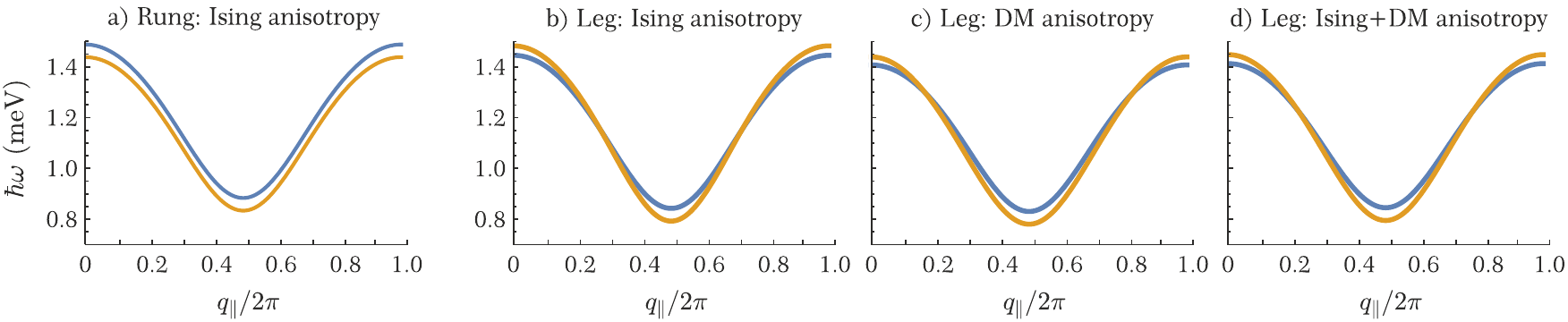}
	\caption{\label{fig:DispAniso} 
		Triplet dispersion for a spin ladder with exchange anisotropy: a) Ising anisotropy on the ladder rungs, b) Ising anisotropy on the legs, c) DM anisotropy on the legs, and d) both Ising and DM anisotropy on the ladder legs, in the ratio as predicted for a simple super-exchange mechanism (Eq.~\ref{eq:SingleBondExchangeAniso}). For these plots, isotropic $J_\bot/k_\mr{B}=12.7$~K and $J_\|/k_\mr{B}=3.54$~K were used, approximately describing the ladders in BPCB. Further, in all four cases the anisotropy parameter was chosen such that the triplet splitting at the band minimum amounts to 50~$\mu$eV. In all plots, the blue line shows the doubly degenerate $\epsilon_\pm(k)$ dispersion and the orange line shows the $\epsilon_0(k)$ triplet. }
\end{figure*}

\subsection{Discussion of experimental results}
The experimental data shown in Fig.~\ref{fig:spectra} closely resembles previously published results on BPCB with dispersive triplet excitations \cite{Savici2009,ThielemannRuegg2009}. In these experiments the resolution was not sufficient to detect the small band splitting and those data were very well described using a model of an isotropic spin ladder with $J_\bot/k_\mr{B}=12.7(1)$~K and $J_\|/k_\mr{B}=3.54(3)$~K\cite{Blosser2018}.
Only with the superior resolution of the present experiment, the small splitting of the triplet band is observed. 

In our data, perpendicular to the one-dimensional ladders, no dispersion is observed within the experimental resolution. This validates BPCB as an exceptionally one-dimensional ladder compound and it corroborates, that the observed band splitting is inherent to the spin ladders and is not related to residual 3D interactions. In addition, inter-ladder interactions were previously estimated on the order of a few 10s of mK\cite{Klanjsec2008NMROrdering} (a few $\mu$eV) -- much smaller than the observed band splitting. 

A previous ESR study on BPCB\cite{CizmarBPCBAniso2010}
found clear signatures of magnetic anisotropy, of exactly the same magnitude as the band splitting observed in the present experiments. 
By a careful study of the angular dependence of the ESR signal, an anisotropy axis tilted approx.~50$^\circ$ from the $b$ axis in the $(b,c^\star)$ plane could be identified as a special direction. However, the origin of this anisotropy remained unclear. 

One source of magnetic anisotropy are dipolar interactions present in all materials\cite{Fazekas1999}. For classical magnetic moments at the shortest Cu--Cu distance found in BPCB, dipolar interactions are estimated as $J_\mr{dipolar}/k_\mr{B}\approx 6$~mK ($< 1$~$\mu$eV), much smaller than the observed band splitting. Dipolar interactions clearly can be ruled out as the primary cause of magnetic anisotropy in BPCB. Since for the $S=1/2$ Cu$^{2+}$ ions there cannot be any single-ion anisotropy, \emph{exchange anisotropy} must be the dominant cause of the observed small splitting of the triplet band. Before further analyzing our data in this regard, we first discuss the nature of anisotropic super-exchange interactions in the following section.

\subsection{Anisotropic super-exchange}
Most generally, the effective exchange interaction between two spins takes the form $\mc{H}_{1,2}=\mb{S}_1 \Gamma \mb{S}_2$ where $\Gamma$ is a tensor. Decomposing $\Gamma$ into its symmetric and anti-symmetric parts and choosing a basis such that the symmetric part is diagonal, this can be written as
\be
\mc{H}_{1,2} = 
J \,\mb{S}_1\!\cdot\mb{S}_2
+ \mb{D}\cdot\left(\mb{S}_1\!\times\mb{S}_2\right)
+ \!\! \sum_{\al=x,y,z}\!\!\! G^\al\, {S}_1^\al {S}_2^\al.
\ee
Here, $J$ is the isotropic (Heisenberg) exchange. The so-called Dzyaloshinskii-Moriya vector $\mb{D}$ quantifies the anti-symmetric exchange and $G^\al$ ($\al=x,y,z$) determine the symmetric (Ising) exchange anisotropy where $\sum_\al G^\al=0$.

Including spin-orbit coupling (SOC) into Anderson's theory of super-exchange \cite{Anderson1959} indeed all these contributions are obtained \cite{Moriya1960,Moriya1960PRL}. However, in this setting, symmetric and anti-symmetric interactions are not independent and for a single bond, to a good approximation the super-exchange interaction Hamiltonian reads \cite{Kaplan1983,SEWA1992,SEWA1993}
\be \label{eq:SingleBondExchangeAniso}
\mc{H}_{1,2} = 
\left(\! J \!-\frac{|\mb{D}|^2}{4J} \right) \! \mb{S}_1\!\cdot \mb{S}_2  
+ \mb{D}\cdot(\mb{S}_1\!\times \mb{S}_2) + \frac{1}{2J} (\mb{D}\cdot \mb{S}_1 )(\mb{D}\cdot \mb{S}_2 ).   
\ee
Here a \emph{single} vector $\mb{D}$ defines both the symmetric and antisymmetric exchange. 
This expression, although not obvious, is fully invariant under spin rotations\cite{SEWA1992,SEWA1993}: While the term $\mb{D}\cdot(\mb{S}_1\times \mb{S}_2)$ acts to confine the spins to a plane perpendicular to the vector $\mb{D}$, the Ising contribution acts to align the spins with $\mb{D}$. Thus, for a single bond, anisotropic super-exchange will never single out a particular direction. 
Nonetheless, for multiple connected bonds, {\it frustration} of the $\mb{D}$ vectors may still beak spin rotation symmetry and lead to the appearance of anisotropy\cite{SEWA1992,SEWA1993}.

Besides this mechanism, considering multiple orbitals of the ligands transmitting the super-exchange interactions, in the presence of SOC may result in {\it additional} sources of exchange anisotropy, beyond Eq.~\ref{eq:SingleBondExchangeAniso}. However, these additional effects have been argued to be much smaller in magnitude\cite{KoshibaeCommentSEWA,ReplySEWA}.

This leaves us with the following situation: When considering super-exchange interactions, to first approximation symmetric and antisymmetric anisotropy contributions always come together pointing in the same direction and at a fixed ratio. In this case, if the center of a bond corresponds to a center of inversion symmetry, not only is antisymmetric exchange anisotropy forbidden by symmetry as is commonly known\cite{DZYALOSHINSKY1958}, but we also expect symmetric exchange anisotropy to vanish. Only additional mechanisms may then lead to Ising-type anisotropy which is in principal allowed on a bond with inversion symmetry.

\subsection{Exchange anisotropy in BPCB}

For the compound BPCB with centers of inversion symmetry on the ladder rungs, we consider the following cases compatible with the crystallographic symmetry:
\begin{itemize}\vspace{-0.3\topsep}
	\setlength{\parskip}{0.5\topsep} \setlength{\itemsep}{0pt plus 1pt}
	\item [a)] Ising anisotropy on the ladder rungs 
	
	\item [b)] Ising anisotropy on the ladder legs \setlength{\parskip}{0pt}
	\item [c)] DM anisotropy on the ladder legs
	\item [d)] Both DM and Ising anisotropy on the ladder legs as predicted for simple super-exchange (Eq.~\ref{eq:SingleBondExchangeAniso}).
\end{itemize}\vspace{-0.3\topsep}
Using the results of Sec.~\ref{sec:TripletDispersion}, for these four cases, the triplet dispersion is plotted in Fig.~\ref{fig:DispAniso}. For these plots, we have used the Heisenberg exchange constants $J_\bot/k_\mr{B}=12.7$~K and $J_\|/k_\mr{B}=3.54$~K approximately describing BPCB\cite{Blosser2018}. In all cases the anisotropy parameter was chosen such that the triplet splitting at the band minimum amounts to $50$~$\mu$eV.

Comparing the dispersion calculated for Ising-type exchange anisotropy on the ladder rungs (Fig.~\ref{fig:DispAniso}a) to the data of Fig.~\ref{fig:spectra}, we can clearly rule out this case as the dominant source of anisotropy in BPCB. The data qualitatively disagree with the calculated dispersion.
For the case of anisotropic leg exchange, all three types of anisotropy considered give qualitatively the same dispersion (Fig.~\ref{fig:DispAniso}b,c,d). There is a doubly degenerate band with a smaller bandwidth and a non-degenerate band with larger bandwidth. In all cases the calculated anisotropy splitting is slightly larger at the band minimum than at the band maximum, just as in our data%
\footnote{In Fig.~\ref{fig:DispAniso} the anisotropy parameter was chosen so as to ensure $\Delta_\mr{min}=50$~$\mu$eV. For the different models of leg anisotropy (Ising only, DM only, both), this gives $\Delta_\mr{max}=37,32$~and~$35$~$\mu$eV. However, since from 2\textsuperscript{nd} to 3\textsuperscript{rd} order perturbation expansion the dispersion still slightly changed, these small differences are probably not reliable enough to use as a criterion to select one of these three cases as the most suitable model describing the compound BPCB.}.

\begin{figure}
	\includegraphics{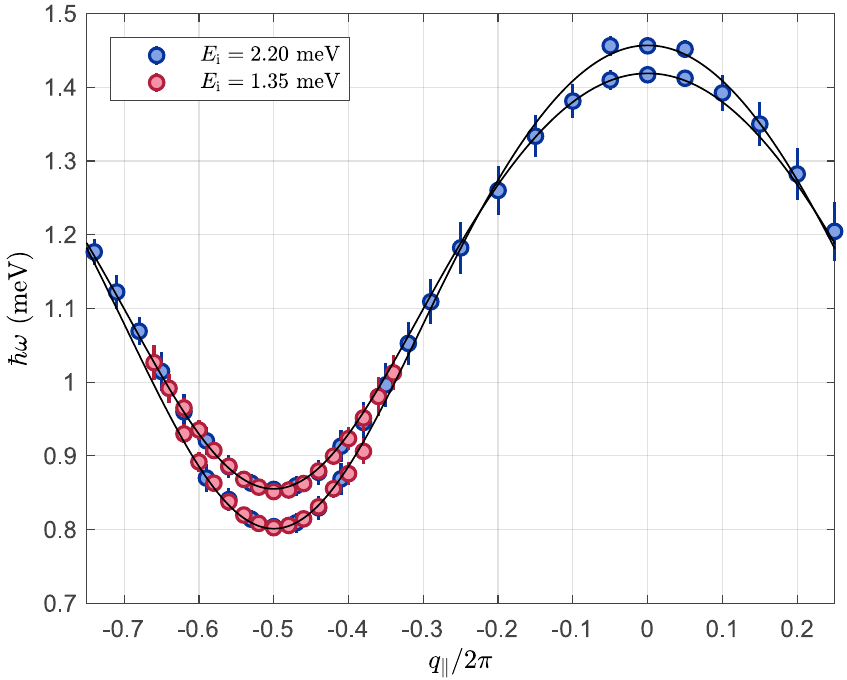}
	\caption{\label{fig:NeutronFit} 
		Position of the triplet bands extracted from the neutron scattering data of Fig.~\ref{fig:spectra}. The dispersion drawn as a solid line represents a model with anisotropic interactions on the ladder legs (case d) as described in the text.
	}
\end{figure}

In Fig.~\ref{fig:NeutronFit} we show the triplet dispersion extracted from our data. These points were obtained from Gaussian fits to constant-$q_\|$-cuts, similar to the ones shown in Fig.~\ref{fig:cuts}.
The vertical bars denote the width of the observed peaks. Fitting the calculated dispersions with anisotropic leg exchange (cases b,c,d) to this data, we find excellent agreement for all three cases.  Indeed, the three calculated dispersions are so similar, that from our data, it is impossible to determine which type of exchange anisotropy on the ladder legs actually causes the observed band splitting in BPCB. 
As an example, in Fig.~\ref{fig:NeutronFit} the solid line shows the dispersion calculated for anisotropic exchange on the ladder legs given by Eq.~\ref{eq:SingleBondExchangeAniso} (case d) 
with fitted parameters
\begin{align}\label{eq:KSEA_FitParameters}
J_\bot/k_B=&12.77(1)\,\mr{K}, \quad \\
J_\|/k_B=3.55(1)\,\mr{K}, &\quad 
D_\|/k_B=1.44(2)\,\mr{K}.\notag
\end{align}
In the parametrization of the Hamiltonian of Sec.~\ref{sec:LadderHamiltonian_LegAniso}, the coefficients are 
$\tilde{J}_\|=J_\| - {D_\|^2}/{(4 J_\|)}$, 
$\lambda = \tilde{J}_\|/{J_\bot}$, 
$A = {D_\|^2}/{  (2 J_\| \tilde{J}_\|) }$ and 
$B = {D_\|}/{\tilde{J}_\|}$.
Here, we stress that the numerical value of $D_\|/k_B=1.44$~K cannot serve as an estimate for the overall magnitude of the magnetic anisotropy which should be  estimated from the band splitting of 50~$\mu$eV, i.e. $\approx0.6$~K in units of Kelvin. 

{
This leaves the question, how one might experimentally determine the type of leg exchange anisotropy dominant in BPCB.
The most promising route to answering this question would be to measure the triplet dispersion as a function of orientation and strength of a small applied magnetic field. In combination with calculations also including an applied magnetic field, such data might indeed allow to determine the precise type of exchange anisotropy on the ladder legs in BPCB. However, this is clearly beyond scope of the present work. 
}

\subsection{Final remarks}
In BPCB the interactions on the ladder rungs are the strongest ones by far ($J_\bot\approx 3.6 \,J_\|$). On these bonds, antisymmetric exchange anisotropy is prohibited by symmetry and only Ising-type exchange anisotropy is allowed. It may thus seem plausible to expect this to be the dominant source of anisotropy.
Yet, experimentally we find that the anisotropy is predominantly due to the much weaker leg interactions of the ladder. Indeed, this is fully consistent with the theory of super-exchange interactions, where symmetric and antisymmetric anisotropy contributions are always coupled. If either is prohibited by symmetry, also the other will vanish. The negligible Ising anisotropy on the ladder rungs in BPCB we thus see as supporting evidence, for the theory of anisotropic super-exchange\cite{SEWA1992,SEWA1993}. 

Since the theory of super-exchange predicts DM anisotropy $B\propto |\mb{D}|$ and Ising anisotropy $A\propto |\mb{D}|^2$, initially the latter was considered subdominant\cite{Moriya1960PRL,Moriya1960}. However, both contributions are now understood to be equally important\cite{Kaplan1983,SEWA1992,SEWA1993}. Whilst not immediately intuitive, the triplet dispersions calculated in Sec.~\ref{sec:LadderHamiltonian_LegAniso} illustrate this effect: In the obtained expressions $A$ appears linearly, while $B$ only appears to second order as $B^2$.

Finally, we mention an apparent contradiction: Super-exchange can never give rise to DM anisotropy exclusively. Yet, oftentimes pure DM interactions are used very successfully to explain anisotropy effects observed experimentally. 
Here, the present case study offers some illustration. For the spin ladder, different types of exchange anisotropy on the ladder legs lead to almost identical triplet dispersions that all explain our data equally well. We speculate that also in other systems one might encounter similar situations and it is for this reason that pure DM anisotropy is used so successfully to explain experimental findings, even though (within the theory of super-exchange) it lacks any microscopic justification. 
We are only aware of one experimental study on the helimagnet \BaCuGeO{}, where a model employing the full Hamiltonian for anisotropic super-exchange interactions (Eq.~\ref{eq:SingleBondExchangeAniso}) was compared to experimental data and indeed better agreement was found than for a model employing DM anisotropy only\cite{ZheludevPRL1998}.

Further case studies of the nature of magnetic exchange anisotropy would certainly be interesting. We also suggest, that whenever a model with pure DM interactions is considered, it would be enlightening to also consider a model with both Ising and DM anisotropy, compatible with super-exchange.

\section{Conclusion}
The prototypical spin ladder compound \BPCB{} (BPCB) has been studied by means of high resolution inelastic neutron scattering. We find a small splitting of the triplet band of $50(1)$~$\mu$eV at the band minimum and $40(2)$~$\mu$eV at the band maximum. 
Further, for a spin ladder with exchange anisotropy, the triplet dispersion is calculated in a strong coupling expansion.

The Ising-type anisotropy allowed by crystallographic symmetry on the ladder rungs (by far the strongest bonds) we find to be negligible, in line with the theory of anisotropic super-exchange.
Three models with exchange anisotropy on the ladder legs, all describe the data equally well. Whilst we cannot distinguish these by comparison to our data, we note that only one is compatible with super-exchange interactions.

{\it Acknowledgement.}
This work is partially supported by the Swiss National Science Foundation under Division II. Experiments at the ISIS Neutron and Muon Source were supported by a beamtime allocation RB1720009 from the Science and Technology Facilities Council, UK.

\appendix*
\section{Strong coupling expansion}
\label{app}
	
In this appendix we give some details on the strong coupling expansion calculation of the triplet dispersion in a strong rung spin ladder following Ref. \onlinecite{ReigrotzkiRice1994}:
First, consider a single Heisenberg $S=1/2$ spin dimer with Hamiltonian $\mc{H}^0=J_\bot\, \mb{S}_{1} \!\cdot \mb{S}_{2}$. The eigenstates are given by a singlet state $\ket{s}$ with energy $E_s=-\frac{3}{4}J_\bot$ and three triplet states $\ket{t^\sigma}$ with energy $E_t=\frac{1}{4}J_\bot$ and $S_z$ spin component $\sigma=-1,0,1$.
In all cases our calculations start from the strong rung limit, where the spin ladder is nothing but an array of $L$ independent Heisenberg spin dimers 
$$
\mc{H}^\bot = J_\bot \sum_{R=1}^{L}  \mb{S}_{R,1} \cdot \mb{S}_{R,2}.
$$
The ground state is a product of singlets $\ket{0}=\ket{s\dots s}$ with energy $E_0=L E_s$. The first excited states are the $3L$ degenerate single triplet states $\ket{R,\sigma}=\ket{s\dots t_R^\sigma \dots s}$ where the $R$\textsuperscript{th} rung is excited to an $S_z=\sigma$ triplet.

Now we introduce the much weaker interaction $\mc{H}'$ as a perturbation. It connects the individual dimers giving the spin ladder Hamiltonian $\mc{H} = J_\bot \left( \mc{H}^\bot + \lambda \mc{H}' \right)$ with $\lambda\ll 1$. 
In section \ref{sec:LadderHamiltonian_Heis}, $\mc{H}'$ describes Heisenberg exchange connecting the dimers via the ladder legs. In addition to this, in Secs. \ref{sec:LadderHamiltonian_RungAniso} and \ref{sec:LadderHamiltonian_LegAniso},  $\mc{H}'$ also contains anisotropic exchange contributions on the ladder rungs or legs, respectively.

The ground state energy in the presence of $\mc{H}'$ we denote as $\tilde{E_0}$. It is computed using standard non-degenerate Rayleigh-Schr\"odinger perturbation theory. The triplet states however are $3L$-fold degenerate and some care is required. The $L$ fold degeneracy is lifted to first order in $\lambda$ and the appropriate eigenstates are Bloch waves
$$
\ket{k,\sigma}=\frac{1}{\sqrt{L}}\sum_{R=1}^{L}e^{i k R}\ket{R,\sigma}
$$
for $k=\frac{n}{2\pi}$, $n=0,\dots,L-1$. In the absence of $\mc{H}'$ they have energy $E_{k,\sigma}=E_0+J_\bot$. Further we note that all $\mc{H}'$ commute with $S^z = \sum_{R=1}^L S_R^z$ and thus with the unperturbed $\mc{H}^\bot$. Therefore, they will never mix triplets with different quantum numbers $\sigma$. Starting with the $\ket{k,\sigma}$ triplet Bloch waves we can therefore again use non-degenerate perturbation theory to obtain their energy $\tilde{E}_{k,\sigma}$ in the presence of the perturbation $\mc{H}'$.
Calculating the ground state energy $\tilde{E}_0$ and the triplet energy $\tilde{E}_{k,\sigma}$ to third order perturbation theory in $\lambda$, we obtained the triplet dispersions $\epsilon_\sigma(k)=\tilde{E}_{k,\sigma}-\tilde{E}_0$. For the different $\mc{H}'$, these are given in Sec. \ref{sec:TripletDispersion}.

\bibliography{referencesBPCB_Aniso}

\end{document}